# Lithium attachment to $C_{60}$ and nitrogen- and boron-doped $C_{60}$: a mechanistic study

Yingqian Chen[a], Chae-Ryong Cho[b], Sergei Manzhos[c,1]

[a] Department of Mechanical Engineering, National University of Singapore, Block EA #07-08, 9 Engineering Drive 1, Singapore 117576

[b] Department of Nanoenergy Engineering, Pusan National University, 63 Beon-gil 2, Busandehak-ro, GeumJeong-gu, Busan 46241, Republic of Korea

[c] Centre Énergie Matériaux Télécommunications, Institut National de la Recherche Scientifique, 1650 boulevard Lionel-Boulet, Varennes QC J3X1S2, Canada

**Abstract**

Fullerene-based materials including $C_{60}$ and doped $C_{60}$ have previously been proposed as anodes for lithium ion batteries. It was also shown earlier that *n*- and *p*-doping of small molecules can substantially increase voltages and specific capacities. Here, we study *ab initio* the attachment of multiple lithium atoms to $C_{60}$, nitrogen-doped $C_{60}$ (*n*-type), and boron doped $C_{60}$ (*p*-type). We relate the observed attachment energies (which determine the voltage) to changes in the electronic structure induced by Li attachment and by doping. We compare results with a GGA functional and a hybrid functional and show that while they agree semi-quantitatively with respect to the expected voltages, there are qualitative differences in the electronic structure. We show that, contrary to small molecules, single atom *n*- and *p*-doping will not lead to practically useful modulation of the voltage-capacity curve beyond the initial stages of lithiation.

---

1  Author to whom correspondence should be addressed. E-mail: Sergei.Manzhos@emt.inrs.ca; Tel: +1 514 2286841.

## Introduction

Fullerene-based materials have been explored for use as active electrode materials for Li and Na ion batteries in several works.[1-5] Cyclic voltammetry of $C_{60}$ in solution identified well-resolved peaks at -1.07, -1.43, and -1.92 V vs Fc/Fc$^+$ in Ref. 6 (about 2.4, 2.1, and 1.6 V, respectively, vs Li/Li$^+$). In Ref. 7, reduction peaks for thin film $C_{60}$ were reported around -0.5, -1.0, -1.7, and -2.2 V vs Fc/Fc$^+$. In Ref. 8, an expanded measurement window was used and reduction potentials ($E_{1/2}$ values) relative to Fc/Fc$^+$ were measured in solution at -0.98, -1.37, -1.87, -2.35, -2.85, and -3.26 V (about 2.4, 2.1, 1.6, 1.1, 0.6, and 0.2 V, respectively, vs Li/Li$^+$). When $C_{60}$ is used as an active electrode material, voltammograms recorded during lithiation show a large sloping plateau in the 0.7 V area and short plateaus around 1.5, 1.9 and 2.3 V vs Li/Li$^+$.[2] Experiments indicate a maximum specific capacity reaching 400 mAh/g, corresponding to insertion of 10 to 12 Li atoms per molecule.[2] The broad plateau in the 0.7 V area extends from $Li_3C_{60}$ to $Li_{10}C_{60}$ (corresponding to a specific capacity of 372 mAh/g) and $Li_{12}C_{60}$ is reached at about 0.1 V.[2] Formation of $Li_{10}C_{60}$ corresponds to $C_{60}$ accepting 10 valence electrons from the Li atoms into its lowest unoccupied molecular orbitals. In this configuration, similar to other known complexes like $K_{10}C_{60}$,[9] all 10 attached Li should be fully ionized.

The voltages observed with $C_{60}$ imply its possible use as an anode. In the anode application, one desires either a voltage close to 0 vs Li/Li$^+$ (while remaining sufficiently positive to avoid plating) or a voltage just above 1.3 V which would avoid reduction of common carbonate based electrolytes and thus enable high-rate operation unhampered by the SEI (solid-electrolyte interphase) and plating/dendrites.[10-12] The main plateau of a $C_{60}$ electrode at around 0.7 V does not ideally achieve either of these purposes for a Li ion battery. It is possible to modulate the voltage-capacity curve by chemical modification including functionalization and doping. Functionalized and doped $C_{60}$ based materials have been explored in experimental and modelling works for use in Li ion batteries. Examples are Ag-doped/functionalized fullerenol[13] or substitutional nitrogen-doped $C_{59}N$ and functionalized $C_{59}N$.[14]

Doping of the active electrode material is a powerful technique to modulate the interaction energy of a semiconductor host with the active cation.[15] In inorganic materials such as monoelemental semiconductors, the initial stage of lithiation, sodiation, magnesiation etc. involves donation of the valence electron to the conduction band of the host;[16-25] in oxides, the valence electron can occupy a state in the conduction band.[26-31] These are relatively high-energy states; $p$-doping can be used to create unoccupied states near the top of the valence band which can be occupied by the valence electron of the alkali atom and thereby strengthen the binding via the bandstructure part of the total energy, which can either increase the voltage or induce electrochemical activity in materials which are inactive in undoped state.[15,16] *n*-Doping, on the other hand, is not expected to improve the voltage is inorganic solids .[15] This is ultimately related to small reorganization energies of such materials. Many organic electrode materials[32] are semiconductors, and the strategy of *p*-doping is fully applicable to them; in Ref. 33, we computed that *p*-doping can lead to strengthening of the binding energy on the order of 2 eV (voltage increase on the order of 2 V) when using typical organic small molecule building blocks. With organic molecules, in contrast to inorganic solid semiconductors, also *n*-doping was shown to lead to stronger binding, this was related in Ref. 33 to strain effects. Here, we will show that in molecules, stronger binding (higher voltages) induced by *n*-doping can also be understood from the bandstructure perspective (in the following, we will liberally use "bandstructure" as set language even when talking about molecules). Substitutions not amounting to doping (where the number of valence electrons does not change) can also be used to strengthen interaction with the active cation; for example, we showed that replacing a -CH- group with N in disodium terephthalate[34,35] to result in disodium pyridine dicarboxylate leads to increased voltage for first Na attachment per formula unit, in agreement with experiment.[36,37] This was related to a bond formation between Na and N.

In small-molecule systems, the theoretical capacity is typically reached with a small number of Li or Na attached per formula unit. Specifically in materials operating by the insertion and reduction mechanism, the specific capacity is typically reached with two Li or Na per molecule, corresponding to full occupancy of the molecular LUMO (lowest unoccupied molecular orbital).[10,36,38] In this case, an increase of the voltage for first Li, Na attachment is *practically* relevant, as it significantly modifies the voltage-capacity curve, at least up to a half of the theoretical capacity. For example, a voltage-capacity curve with two plateaus was obtained in Ref. 36 which decreased segregation into the fully sodiated state and improved cycle rate and life. In inorganic hosts, to induce practically relevant improvement in the voltage-capacity curve, a high dopant concentration is needed, to the tune of several at%, which is experimentally feasible for some materials.[17,39] One may therefore question whether a single dopant can substantially change the voltage-capacity curve of $C_{60}$[14] considering that the final state of charge involves at least 10 Li atoms per molecule.

Density functional theory (DFT) models are able to describe experimental voltage-capacity curves with semi-quantitative accuracy.[40] Typically, this requires periodic solid-state calculations, which limits the range of practically applicable approximations. Specifically, it is more difficult and CPU-costly to use hybrid functionals, and GGA functionals remain the most widely used in this application. This, even though the charge transfer nature of the Li – host material interactions makes desirable the inclusion of exact exchange. Molecular or oligomeric models can be used for organic materials.[32] For materials operating by oxidation, molecular / oligomeric models are sometimes able to predict the voltage capacity curve.[41] For materials operating by insertion / reduction, we observed in a series of studies that the qualitative features of the voltage-capacity curve, including its shape and any effects of doping are well reproduced in a molecular model, while the absolute magnitude of the voltage is underestimated on the order of 1 V due to the neglect of aggregate state effects.[10,34-38,42] Another disadvantage of a molecular model is artificial persistence of the voltage-capacity curve beyond the theoretical capacity;[34,37,42] we are, however, able to circumvent this issue (see Methods). A significant advantage of a molecular model is ease of application of a hybrid functional (and of wavefunction based methods, as needed) and of comparison between different computational approaches.[41-44] This is also the route taken in this work. A recent DFT based work on N-doped $C_{60}$ also used a molecular model with a hybrid functional.[14]

In this work, we perform an *ab initio* study of Li attachment to $C_{60}$, N-doped $C_{60}$, and B-doped $C_{60}$, the last two chosen as examples of substitutional *n*- and *p*-doping, respectively. We study whether doping can be used to modulate in a practically significant way the voltage-capacity curve of $C_{60}$. We focus on the mechanistic understanding of $Li_n$-$C_{60}$ interaction, derived from the electronic structure; we therefore work with a molecular model which is expected to underestimate the magnitude of the voltage but preserve the shape of the voltage-capacity curve[36,37] and correctly predict the effects due to doping.[15,32] This model allows us to compare the results obtained with a GGA functional and a hybrid functional and thereby assess whether a GGA functional – still most practical for solid state modelling – correctly describes properties of this system such as bandstructure changes during Li attachment and the voltage-capacity curve. In contrast to the model of Ref. 14 which computed the reduction potential and therefore provided a preview into the effects of doping and functionalization on the open-circuit voltage, i.e. the initial part of the voltage-capacity curve (which, from what is known about its shape for $C_{60}$,[2] is not much practically relevant), we consider the effects of doping on the entire curve and show that, contrary to small molecules, single atom *n*- and *p*-doping will not lead to practically relevant modulation of the voltage-capacity curve beyond the initial stages of lithiation.

## Methods

Density functional theory calculations were performed in Gaussian 09[45] using the B3LYP[46] and PBE[47] exchange correlation functionals and the 6-31+g(d,p) basis set. Spin polarization was used for systems with odd numbers of electrons. Tight convergence criteria were used for structure optimization. Initial $C_{59}N$, $C_{59}B$ structures were obtained by replacing one C atom with an N or B atom, respectively. Bader charges were computed with the Bader analysis program.[48] Partial densities of states (PDOS) were produced with the GaussSum program,[49] and visualizations with VESTA.[50] Bond formation was identified by using charge density differences, $\Delta\rho = \rho_{Li-sys} - (\rho_{Li} + \rho_{sys})$. where where *sys* is one of $C_{60}/C_{59}N/C_{59}B$, and electron densities $\rho$ for Li and *sys* are computed at the geometries of Li-*sys*. Formation energies $E_f$ of complexes $Li_n$-$C_{60}/C_{59}N/C_{59}B$ were computed as

$$E_f = E_{nLi-sys} - nE_{Li_{bcc}} - E_{sys} \quad , \tag{1}$$

where $E_{nLi-sys}$ is the energy of the complex $Li_n$- $C_{60}/C_{59}N/C_{59}B$ and $Li_{bcc}$ is an estimate of the energy of one Li atom in a bcc (body-centered cubic) structure, which is computed as $Li_{bcc} = E_{atom} - E_{coh}$, where $E_{atom}$ is the energy of a Li atom computed in Gaussian 09 and $E_{coh}$ is the cohesive energy of bcc Li taken as 1.63[51,52] eV/atom. Voltage-capacity curves were computed from piece-wise voltages $V$ between Li fractions $n_1$ and $n_2$ as[40]

$$V = -\frac{E_{n_2 Li-sys} - E_{n_1 Li-sys} - (n_2 - n_1) E_{Li_{bcc}}}{n_2 - n_1} \quad , \tag{2}$$

where $n_1$ and $n_2$ correspond to inflection points on the convex hull built from the dependence $E_f(n)$. Multiple Li configurations (attachment sites) were tried, and the lowest formation energies were used to build the convex hull and to compute $V$.

As discussed in the Introduction, the absence of aggregate state effects not only causes a shift of the voltage-capacity curve as a whole but also makes it less obvious at what state of charge the maximum capacity would be reached. We use the extent of Li ionization – more specifically an abrupt change of its degree – as an indicator of the maximum capacity expected in an experiment. Indeed, we noted in our previous comparative studies of molecular and solid materials that while in molecular calculations the capacity persists beyond that computed in solid state (and that observed experimentally), those states of charge correspond to a markedly lower degree of Li ionization and to occupancy of Li-centred states.[10,38,42]

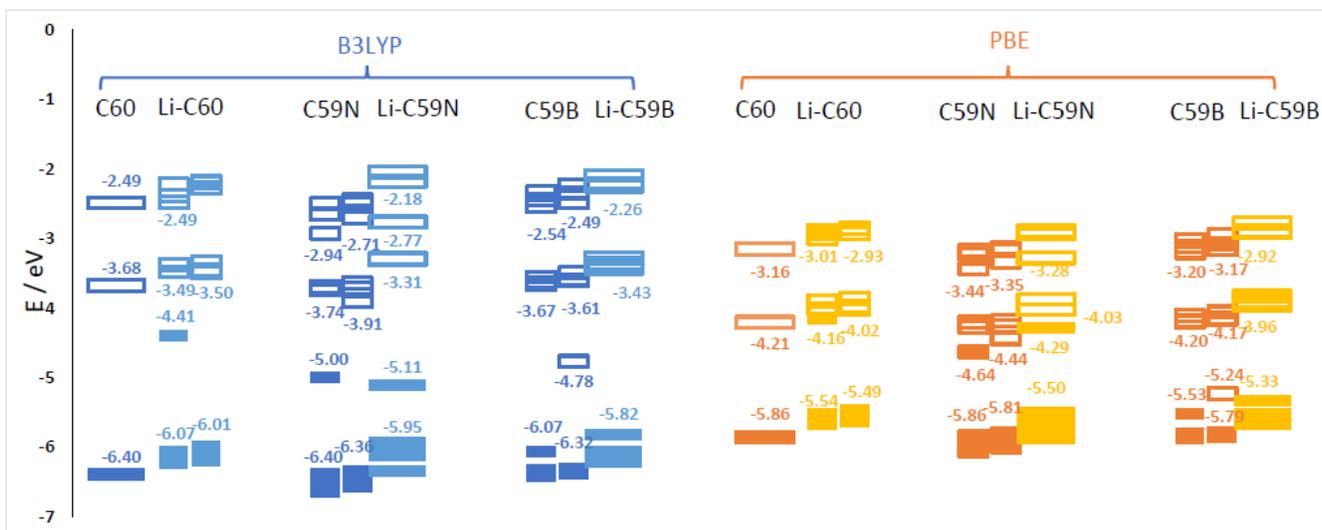

**Fig. 1**. Energies of selected molecular orbitals (MOs) of $C_{60}$, $C_{59}N$, $C_{59}B$ and corresponding molecules with one Li attachment computed with B3LYP and PBE functionals. The filled symbols denote occupied MOs and the empty ones unoccupied MOs. For open-shell cases, both alpha (left) and beta (right) spin channel MOs are shown.

## Results

**Bandstructure of Li-$C_{60}$/$C_{59}$N/$C_{59}$B complexes**

Energies of relevant molecular orbitals of $C_{60}$, $C_{59}N$, $C_{59}B$ and their complexes with Li are shown in **Fig. 1**. We note that the orbital corresponding to the valence electron of a free Li atom is $E$(Li) = -3.65 eV with B3LYP and -3.22 with PBE, i.e. higher than the LUMO of $C_{60}$ by 0.03 eV with B3LYP and by 0.99 eV with PBE. For spin-polarized systems, both spin channels are shown. Li donates its valence electron to the LUMO of $C_{60}$ (see **Fig. 2**). While PBE results in an artificially contracted HOMO (highest occupied molecule orbital) – LUMO gap, this phenomenon of Li ionization is qualitatively similar with both functionals. Upon charge donation, the resulting SOMO (single occupied molecular orbital) is stabilized vs the original LUMO by 0.73 eV with B3LYP but is *destabilized* by 0.05 eV with PBE; as a result the band energy of this electron is different by only 0.25 eV between the functionals, while with both functionals, the $C_{60}$'s HOMO is destabilized by a similar amount (about 0.35 eV). The stabilization of the band energy of Li valence electron upon donation also differs by only 0.2 eV between the functionals (and is stronger with PBE). This helps explain the relatively small (given major quantitative and qualitative differences in bandstructure) difference in voltages computed with the two functionals (on the order of 0.2 V, *vide infra*).

The effect of the substitutional N dopant is to introduce an occupied state (SOMO) in the gap in one spin channel. The SUMO (single unoccupied molecular orbital) of the other spin channel is occupied by the Li valence electron upon Li attachment and stabilized by 1.2 eV, from -3.91 eV to -5.11 eV (forming the HOMO of the complex shown in **Fig. 2**), with B3LYP but is *destabilized* by 0.15 eV, from -4.44 eV to -4.29 eV, with PBE. The bandstructure energy of the valence electron of Li is therefore stabilized by 1.46 and 1.07 eV upon attachment to $C_{60}$ with B3LYP and PBE, respectively. The destabilization of the molecular HOMO (HOMO-1 of the complex) is of similar magnitude with both functionals. The bandstructure argument implies a stronger effect *of the doping* on the open-circuit voltage with B3LYP; this is indeed confirmed by the computed voltage-capacity curve below, although to a smaller degree than suggested by the bandstructure. However, is this case the bandstructure argument is less directly applicable than in the case of pristine $C_{60}$; the bond Li-$C_{59}$N is less ionic, and we can observe covalent bond formation between Li and N, as shown in **Fig. 2**. This is similar to Na-N covalent bond formation was previously observed in disodium pyridine dicarboxylate.[36,37] A critical difference between B3LYP and PBE in this case is due to the fact that Li attachment in this model corresponds to going from a state with 2 unpaired electrons on each component (well-separated Li and $C_{59}$N) with negligible exchange energy to a singlet state of $C_{59}$N stabilized by contributions from exact exchange in the case of B3LYP but not PBE. We therefore expect a stronger open-circuit voltage increase with B3LYP than with PBE, which is indeed what we compute below with Eq. 2.

In the case of boron doping, substitutional B leads to the appearance of an unoccupied state in the gap and of a SOMO derived from a half-occupied C60 HOMO. Upon Li attachment, this state is strongly stabilized forming a doubly occupied C60 HOMO-like orbital. Valence electron of Li thus occupies a state by 2.17 eV lower vs isolated atom with B3LYP and by 2.11 eV lower with PBE. The effects of B doping on the open-circuit voltage are expected to be strong and of similar magnitude with both functionals, by the valence bandstructure argument. We indeed find below that they are strong and similar albeit smaller than 2 V due to electron correlation and other effects. The different effect of the hybrid vs GGA functional here is explained by the fact that with *n*-doping by nitrogen, the energy of the orbital occupied by the Li valence electron is in the gap and much above the C60 HOMO and can be stabilized by exact exchange contributions, while with *p*-doping by boron, it is near C60 HOMO with a relatively high density of states (HOMO to HOMO-4 of C60 have similar energies shown as one merged bar in **Fig. 1**) and likely cannot be stabilized further than the top of the valence band due to electron correlation effects.

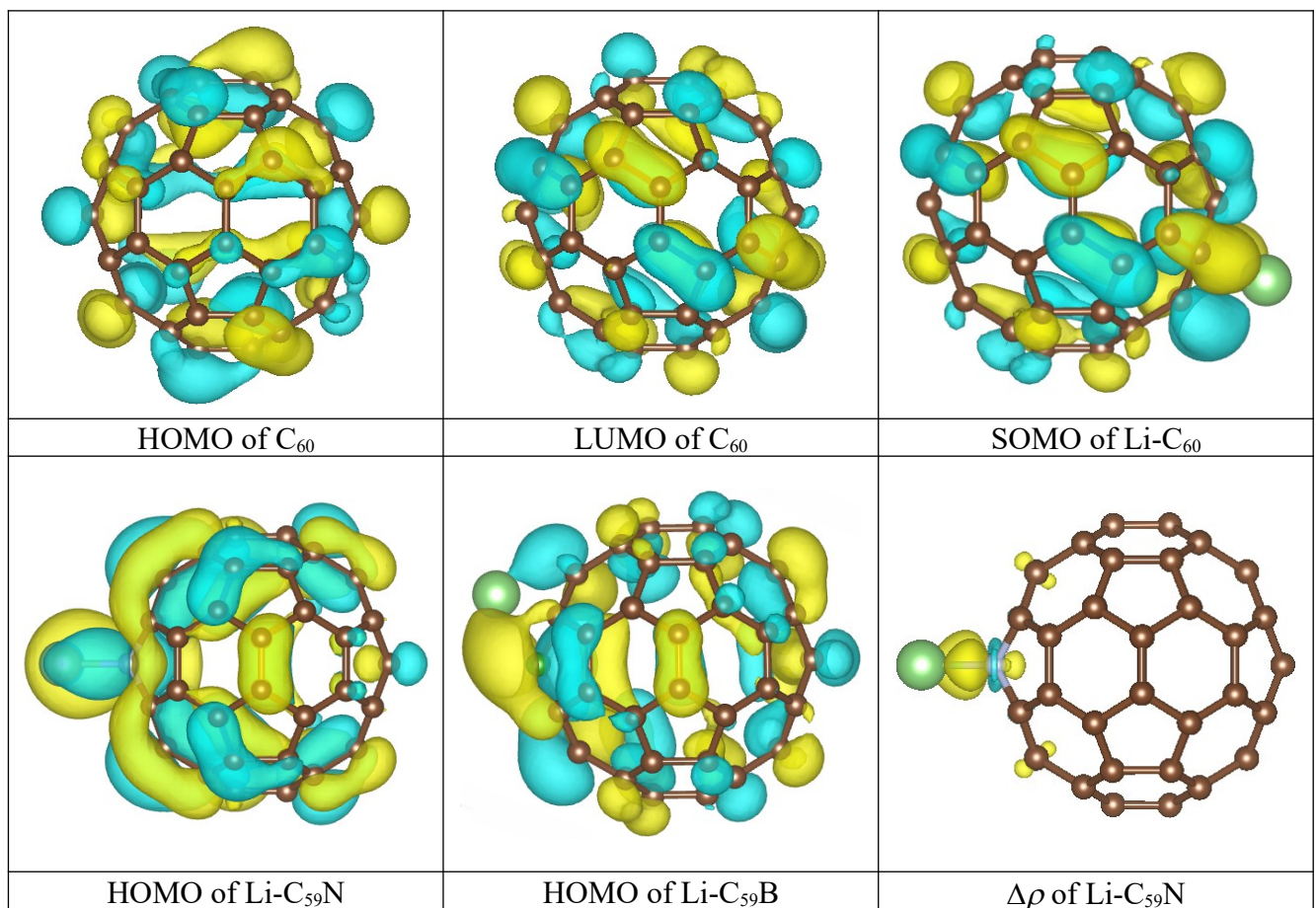

**Fig. 2**. HOMO, LUMO of $C_{60}$ and the orbital occupied by the Li valence electron in Li-$C_{60}$, Li-$C_{59}$N, and Li-$C_{59}$B, as well as charge density difference map $\Delta\rho$ showing bond formation between Li and N dopant. Results of B3LYP calculations are shown, those with PBE are visually similar. Isosurface values are 0.02 $e^{1/2}$/Å$^3$ for orbitals and 0.005 $e^{1/2}$/Å$^3$ for the charge density difference map. The atomic color scheme is C-brown, N-grey, B-green, Li-light green.

**Formation energies and voltage-capacity curves**

The formation energies of the lowest energy Li$_n$-$C_{60}$/$C_{59}$N/$C_{59}$B complexes are shown in **Fig. 3**. The corresponding voltage-capacity curves are shown in **Fig. 4**. The curves are down-shifted vs experiment[2] on the order of 1 V due to the neglect of effects due to molecular packing in a solid.[10,36-38,42] For undoped $C_{60}$, B3LYP predicts an almost flat curve (plateaus at -0.19 V up to Li$_4$C$_{60}$ and -0.22 V thereafter) at around -0.2 V up to Li$_{12}$C$_{60}$. PBE predicts a plateau near 0 up to Li$_3$C$_{60}$ followed by a plateau at -0.13 V up to Li$_{12}$C$_{60}$, suggesting a theoretical capacity of 446 mAh/g. The two functionals, in spite of resulting in significantly different electronic structures, end up predicting voltage-capacity curves differing only on the order of 0.1 V. This is rationalized in the proceeding section.

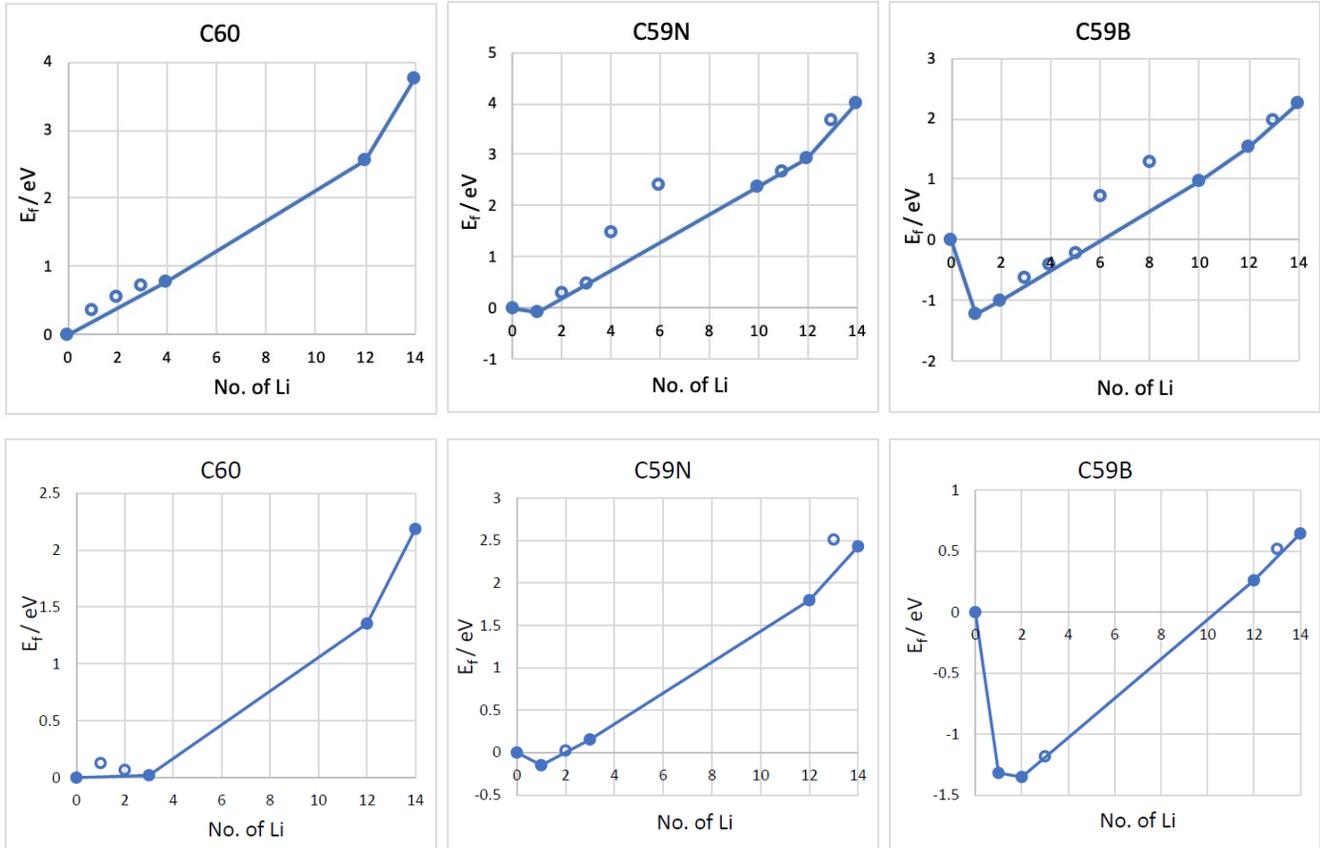

**Fig. 3**. Formation energies ($E_f$) vs the number of Li atoms attached to the $C_{60}$, $C_{59}N$, and $C_{59}B$ molecules computed with B3LYP (top panels) and PBE (bottom panels) functionals. The most stable $E_f$ for each concentration are plotted. The line connects the stable phases (filled points) during Li attachment, forming a convex hull.

The final state of charge with $n = 12$ is in decent agreement with the experiments of Ref. 2. The computed curve shows a drastic drop after that point which corresponds to a sudden change in the degree of ionization of Li. We use the degree of ionization to call the final state of charge expected in an experiment. **Table I** lists Bader charges on Li in $Li_n$-$C_{60}$/$C_{59}N$/$C_{59}B$. For attachment up to a dozen Li per $C_{60}$, all Li atoms are practically fully ionized, with Bader charges on the order of 0.9 $|e|$ for all Li. Beyond $n = 12$, some Li atoms show a markedly lower degree of ionization, this point correspond to a down-step in the voltage-capacity curve in **Fig. 4**. This behaviour is similar with B3LYP and PBE. This is therefore the computed final state of charge. With N doping, the drop in the degree of ionization starts a little earlier and with B doping a little later, as is expected with an electron rich *n*-doped system and an electron-deficient *p*-doped system, respectively. Eq. 2 predicts a similar theoretical capacity for $C_{59}N$ as for $C_{60}$ but a slightly larger capacity for $C_{59}B$ (**Fig. 4**).

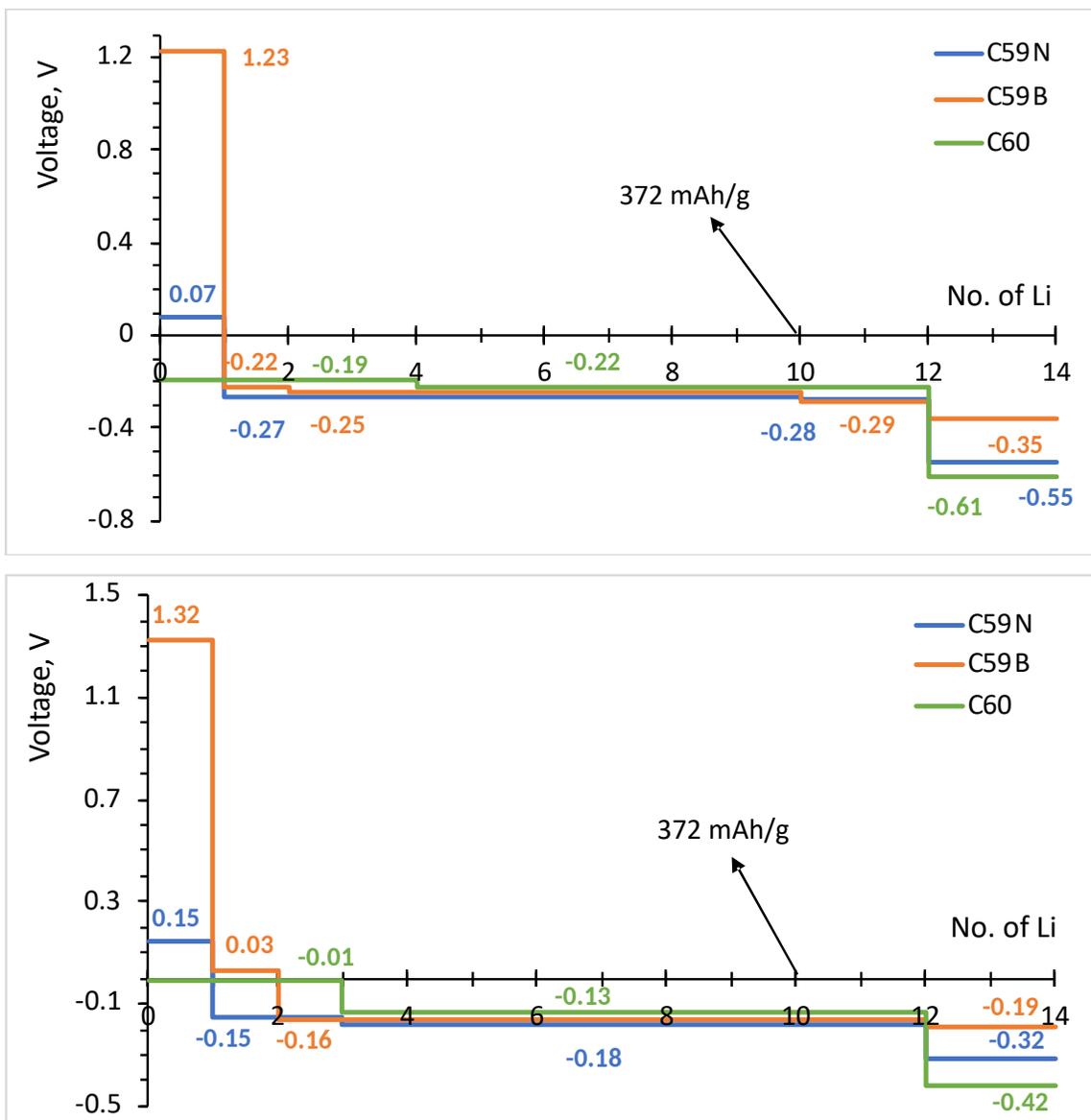

**Fig. 4**. Voltage-capacity curves estimated for lithium attachment to molecular $C_{60}$, $C_{59}N$ and $C_{59}B$ computed with B3LYP (top) and PBE (bottom) functionals.

Table I. Bader charges on Li atoms for attachment of different numbers of Li atoms to $C_{60}$, $C_{59}N$, and $C_{59}B$ computed with B3LYP and PBE functionals. Average charge per Li is given when the charges are similar among Li atoms, and the charge of the last Li attached is shown after "/" when it is significantly different from the average of charges of all other Li atoms.

| Bader charge, $|e|$ | B3LYP | | | PBE | | |
|---|---|---|---|---|---|---|
| | $C_{60}$ | $C_{59}N$ | $C_{59}B$ | $C_{60}$ | $C_{59}N$ | $C_{59}B$ |
| 1 Li | 0.91 | 0.90 | 0.90 | 0.90 | 0.90 | 0.90 |
| 2 Li | 0.90 | 0.90 | 0.90 | 0.90 | 0.90 | 0.89 |
| 3 Li | 0.90 | 0.89 | 0.89 | 0.90 | 0.90 | 0.89 |
| ... | ... | ... | ... | ... | ... | ... |
| 12 Li | 0.86 | 0.85 / 0.34 | 0.86 | 0.86 | 0.84 / 0.45 | 0.86 |
| 14 Li | 0.84 / 0.59 | | 0.85 / 0.74 | 0.84 / 0.62 | | 0.85 / 0.74 |

The attachment of multiple Li atoms to $C_{60}/C_{59}N/C_{59}B$ corresponds to occupation of a number of unoccupied molecular states equal to the number of Li atoms, as is seen from the PDOS plots shown in **Fig. 5**. While for attachment of first Li atoms, there are no significant contributions from Li to the occupied states, for attachment of multiple Li atoms, such contributions are notable, indicating a significant degree of hybridization. Most importantly, any effect of doping is limited to the initial stage of lithiation and has negligible effect on the rest of the voltage-capacity curve responsible for most of the useful specific capacity. There is a strong increase of the voltage for first Li attachment, on the order of 1.5 V for boron doping, with both functionals, as expected from the bandstructure. A smaller increase, on the order of 0.3 V with both functionals, is computed for nitrogen doping. There is a minor effect for the second attached lithium, and no significant effect on the bulk of the plateau extending to *n* = 12.

**Conclusions**

We performed a DFT analysis of storage of multiple Li atoms at $C_{60}$ and *n*- and *p*-doped $C_{60}$ ($C_{59}N$ and $C_{59}B$, respectively). We used a molecular model which allowed us to compare the mechanism of Li storage, i.e. electronic structure changes induced by lithiation, between a hybrid (B3LYP) and a GGA (GGA) functional and among C60, $C_{59}N$, and $C_{59}B$. While the measured voltage-capacity curve for undoped $C_{60}$ is available in the literature,[2] we predict the overall shape of the voltage capacity curve expected in an experiment with $C_{59}N$ and $C_{59}B$. We predict that a moderate increase in the voltage with $C_{59}N$ and a strong (>1 V) increase with $C_{59}B$ vs undoped C60 will be observed for the initial part of the voltage-capacity curve *only*. We predict that no significant changes will be observed for the bulk of the voltage-capacity curve responsible for most of the reversible capacity. We predict a slight increase of theoretical capacity (on the order of 15%) with B doping. That is, contrary to small molecules, single atom *n*- and *p*-doping of C60 will *not* lead to practically useful modulation of the voltage-capacity curve.

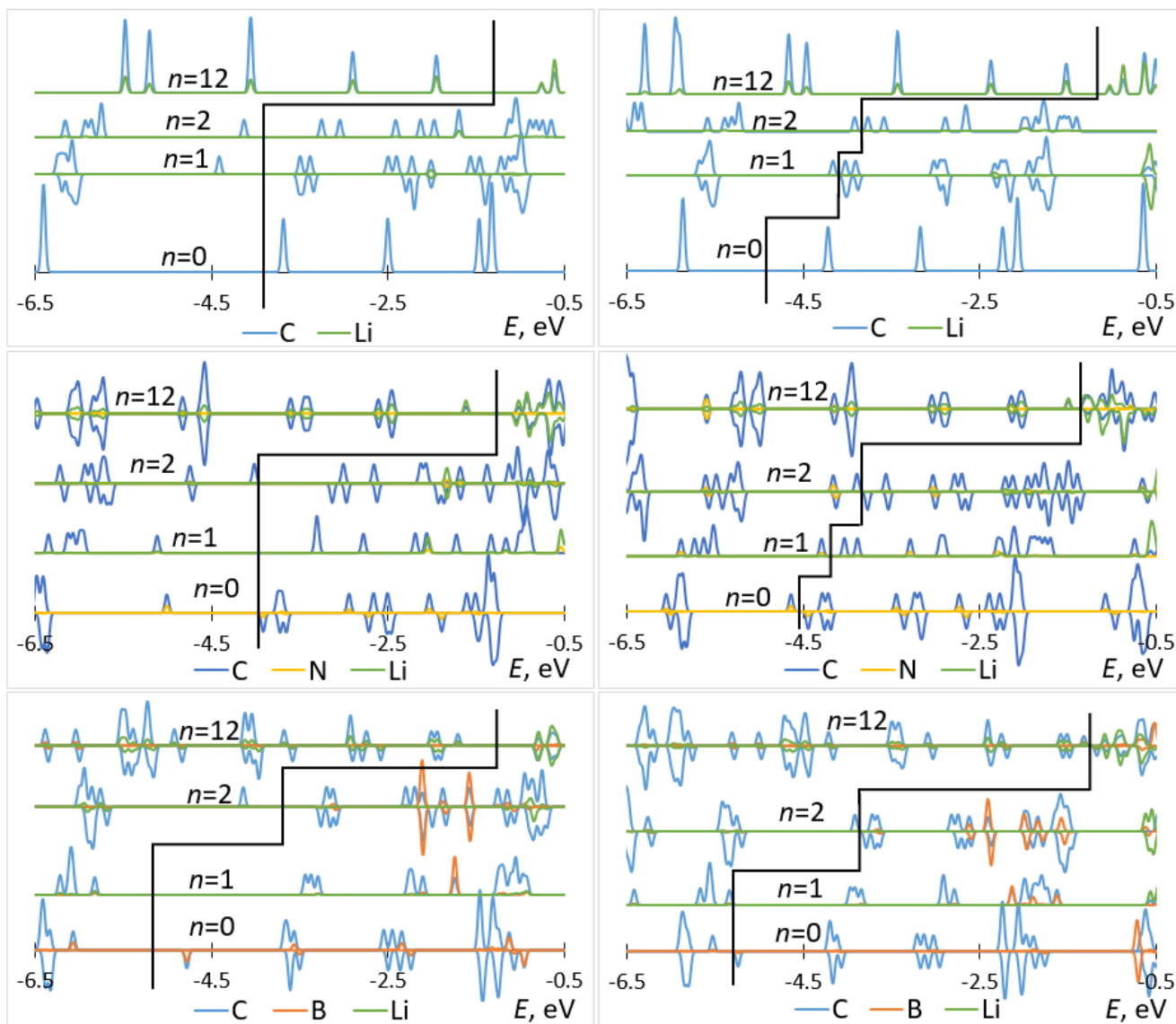

**Fig. 5**. Partial densities of states of Li$_n$-C$_{60}$ (top panels), Li$_n$-C$_{59}$N (middle panels), and Li$_n$-C$_{59}$B (bottom panels) computed with B3LYP (left panels) and PBE (right panels) for selected values of *n*. The black elbow line is used to mark off occupied (left) and unoccupied (right) states.

We observed differences in the electronic structure changes due to Li attachment between B3LYP and PBE functionals, and these changes were also different between *n*- and *p*-doping.

Specifically, the exact exchange contribution much stabilizes the gap state formed by Li $s$ electron donation to the SUMO of $C_{59}N$, while with PBE this state is slightly destabilized, leading to a stronger bump to the open-circuit voltage with B3LYP. In contrast, with *p*-doping, the orbital at the top of the valence band occupied by the Li electron does not benefit from the extra stabilization, leading to a similar in magnitude bump to the open-circuit voltage with both functionals.

B3LYP and PBE functionals, in spite of resulting in significantly different electronic structures, end up predicting voltage-capacity curves differing only on the order of 0.1 V. This was rationalized based on bandstructure changes induced by lithiation. On one hand, this result is good news for solid-state modeling where PBE functionals remain the most practical solution, even though the charge transfer nature of the Li – host material interactions makes desirable the inclusion of exact exchange; on the other hand, our results also show that this apparent similarity in voltages hides significant (quantitative and qualitative) differences in electronic structure changes induced by Li attachment. Important mechanistic details could therefore be missed even with a quantitatively accurate voltage curve.

The presence of the dopant atoms is expected to significantly increase the voltage only up to states of charge where the number of attached Li is just sufficient to fill the empty states created by the dopants.[1] We therefore expect that even with fullerenes with a higher heteroatom content such as $C_{57}N_3$[53] only a small part of the voltage-capacity curve will be modulated. The present model considered that $C_{60}$ and doped $C_{60}$ molecules would largely preserve their structure and electronic properties in solid state, which is true for vdW crystals of molecular fullerenes.[54,55] In the future, however, electrochemically induced dimerization of $C_{60}$ and dimerization of $C_{59}N$, which are known to occur, should also be explored.[56,57]

## Conflicts of interest

There are no conflicts of interest to declare.

## Acknowledgements

We thank Dr. Il Jeon of the University of Tokyo for discussions.

## References

1 S. Kawabe, T. Kawai, R. Sugimoto, E. Yagasaki, K. Yoshino, Electrochemical properties of fullerene derivative polymers as electrode materials, *Jpn. J. Appl. Phys.*, 1997, **36**, L1055-L1058.

2 Y. Chabre, D. Djurado, M. Armand, W. R. Romanow, N. Coustel, J. P. McCauley Jr., J. E. Fischer, A. B. Smith III, Electrochemical intercalation of lithium into solid C60, *J. Am. Chem. Soc.*, 1992, **114**, 764-766.

3 S. Lemont, J. Ghanbaja, D. Billaud, Electrochemical intercalation of sodium ions into fullerene, *Mater. Res. Bull.*, 1994, **29**, 465-472.

4 T. Yildirim, O. Zhou, J. E. Fischer, N. Bykovetz, R. A. Strongin, M. A. Cichy, A. B. Smith III, C. L. Lin, R. Jelinek, Intercalation of sodium heteroclusters into the $C_{60}$ lattice, *Nature*, 1992, **360**, 568–571.

5 S. Scaravonati, G. Magnani, M. Gaboardi, G. Allodi, M. Ricco, D. Pontiroli, Electrochemical intercalation of fullerene and hydrofullerene with sodium, *Carbon*, 2018, **130**, 11-18.

6 C. Jehoulet, A. J. Bard, F. Wudl, Electrochemical reduction and oxidation of C60 films, *J. Am. Chem. Soc.*, 1991, **113**, 5456-5457.

7 C. Jehoulet, Y. S. Obeng, Y. T. Kim, F. Zhou, A. J. Bard, Electrochemistry and Langmuir trough studies of fullerene C60 and C70 films, *J. Am. Chem. Soc.*, 1992, **114**, 4237-4247.

8 Q. Xie, Ed. Perez-Cordero, L. Echegoyen, Electrochemical detection of $C_{60}^{6-}$ and $C_{70}^{6-}$: Enhanced stability of fullerides in solution, *J. Am. Chem. Soc.*, 1992, **114**, 3978-3980.

9 R. Tycko, G. Dabbagh, M. J. Rosseinsky, D. W. Murphy, R. M. Fleming, A. P. Ramirez, J. C. Tully, $^{13}$C NMR spectroscopy of $K_xC_{60}$: Phase separation, molecular dynamics, and metallic properties, *Science*, 1991, **253**, 884-886 (1991).

10 Y. Chen, S. Manzhos, Lithium and sodium storage on tetracyanoethylene (TCNE) and TCNE-(doped)-graphene complexes: a computational study, *Mater. Chem. Phys.*, 2015, **156**, 180-187.

11 M. Madian, A. Eychmüller, L. Giebeler, Current advances in $TiO_2$-based nanostructure electrodes for high performance lithium ion batteries, *Batteries*, 2018, **4**, 7.

12 K. Saravanan, K. Ananthanarayanan, P. Balaya, Mesoporous $TiO_2$ with high packing density for superior lithium storage, *Energy Environ. Sci.*, 2010, **3**, 939-948.

13 F. F. Wang, C. Wang, R. Q. Liu, D. Tian, N. Li, Experimental study on the preparation of Ag nanoparticle doped fullerenol for lithium ion battery application, *J. Phys. Chem. C*, 2012, **116**, 10461−10467.

14 P. Sood, K. C. Kim, S. S. Jang, Electrochemical and electronic properties of nitrogen doped fullerene and its derivatives for lithium-ion battery applications, *J. Energy Chem.*, 2018, **27**, 528–534.

15 J. Lueder, F. Legrain, Y. Chen, S. Manzhos, Doping of active electrode materials for electrochemical batteries: an electronic structure perspective, *MRS Commun.*, 2017, **7**, 523-540.

16 F. Legrain, S. Manzhos, A first-principles comparative study of lithium, sodium, and magnesium storage in pure and gallium-doped germanium: competition between interstitial and substitutional sites, *J. Chem. Phys.*, 2017, **146**, 034706.

17 F. Legrain, S. Manzhos, Aluminum doping improves the energetics of lithium, sodium, and magnesium storage in silicon: a first-principles study, *J. Power Sources*, 2015, **274**, 65-70.

18 F. Legrain, O. Malyi, S. Manzhos, Comparative computational study of the energetics of Li, Na, and Mg storage in amorphous and crystalline silicon, *Comput. Mater. Sci.*, 2014, **94**, 214-217.

19 F. Legrain, O. I. Malyi, S. Manzhos, Comparative computational study of the diffusion of Li, Na, and Mg in silicon including the effect of vibrations, *Solid State Ionics*, 2013, **253**, 157-163.

20 F. Legrain, O. I. Malyi, T. L. Tan, S. Manzhos, Computational study of Mg insertion into amorphous silicon: advantageous energetics over crystalline silicon for Mg storage, *MRS Proc.*, 2013, **1540**, mrss13-1540-e03-06.

21 T. L. Tan, O. I. Malyi, F. Legrain, S. Manzhos, Role of inter-dopant interactions on the diffusion of Li and Na atoms in bulk Si anodes, *MRS Proc.* 2013, **1541**, mrss13-1541-f06-13.

22 O. I. Malyi, T. L. Tan, S. Manzhos, A computational study of the insertion of Li, Na, and Mg atoms into Si(111) nanosheets, *Nano Energy*, 2013, **2**, 1149-1157.

23 O. I. Malyi, T. L. Tan, S. Manzhos, A comparative computational study of structures, diffusion, and dopant interactions between Li and Na insertion into Si, *Appl. Phys. Express*, 2013, **6**, 027301.

24 O. I. Malyi, T. L. Tan, S. Manzhos, In search of high performance anode materials for Mg batteries: computational studies of Mg in Ge, Si, and Sn, *J. Power Sources*, 2013, **233**, 341-345.

25 V. V. Kulish, O. I. Malyi, M.-F. Ng, Z. Chen, S. Manzhos, P. Wu, Controlling Na diffusion by rational design of Si-based layered architectures, *Phys. Chem. Chem. Phys.*, 2014, **16**, 4260-4267.

26 V. Kulish, D. Koch, S. Manzhos, Insertion of mono- vs. bi- vs. trivalent atoms in prospective active electrode materials for electrochemical batteries: an ab initio perspective, *Energies*, 2017, **10**, 2061.

27 F. Legrain, O. I. Malyi, S. Manzhos, Insertion energetics of lithium, sodium, and magnesium in crystalline and amorphous titanium dioxide: a comparative first-principles study, *J. Power Sources*, 2015, **278**, 197-202.

28 D. Koch, V. Kulish, S. Manzhos, A first-principles study of the potassium insertion in crystalline vanadium oxide phases as possible potassium-ion battery cathode materials, *MRS Commun.*, 2017, **7**, 819-825.

29 V. Kulish, S. Manzhos, Comparison of Li, Na, Mg and Al-ion insertion in vanadium pentoxides and vanadium dioxides, *RSC Adv.*, 2017, **7**, 18643.

30 D. Koch, S. Manzhos, A comparative first-principles study of lithium, sodium and magnesium insertion energetics in brookite titanium dioxide, *MRS Adv.*, 2019, **4**, 837-842.

31 D. Koch, S. Manzhos, First-principles study of the calcium insertion in layered and non-layered phases of vanadia, *MRS Adv.*, 2018, **3**, 3507-3512.

32 S. Manzhos, Organic electrode materials for lithium and post-lithium batteries: an *ab initio* perspective on design, *Curr. Opin. Green Sustain. Chem.*, 2019, **17**, 8-14.

33 J. Lueder, M. H. Cheow, S. Manzhos, Understanding doping strategies in the design of organic electrode materials for Li and Na ion batteries: an electronic structure perspective, *Phys. Chem. Chem. Phys.*, 2017, **19**, 13195-13209.

34 Sk Mahasin Alam, S. Manzhos, Sodium interaction with disodium terephthalate molecule: an ab initio study, *MRS Adv.*, 2016, **1**, 3579-3584.

35 Sk Mahasin Alam, S. Manzhos, Exploring the sodium storage mechanism in disodium terephthalate as anode for organic battery using density-functional theory calculations, *J. Power Sources*, 2016, **324**, 572-581.

36 H. Padhy, Y. Chen, J. Lüder, S. R. Gajella, S. Manzhos, Palani Balaya, Charge and discharge processes and sodium storage in disodium pyridine-2,5-dicarboxylate anode - insights from experiments and theory, *Adv. Energy Mater.*, 2018, **8**, 1701572.

37 Y. Chen, J. Lueder, S. Manzhos, Disodium pyridine dicarboxylate vs disodium terephthalate as anode materials for organic Na ion batteries: effect of molecular structure on voltage from the molecular modeling perspective, *MRS Adv.*, 2017, **2**, 3231-3235.

38 Y. Chen, S. Manzhos, Comparative computational study of lithium and sodium insertion in van der Waals and covalent tetracyanoethylene (TCNE) -based crystals as promising materials for organic lithium and sodium ion batteries, *Phys. Chem. Chem. Phys.*, 2016, **18**, 8874-8880.

39 O. Moutanabbir, D. Isheim, H. Blumtritt, S. Senz, E. Pippel, D. N. Seidman, Colossal injection of catalyst atoms into silicon nanowires, *Nature*, 2013, **496**, 78-82.

40 A. Urban, D.-H. Seo G. Ceder, Computational understanding of Li-ion batteries, *npj Comput. Mater.*, 2016, **2**, 16002.

41 Y. Chen, S. Manzhos, Voltage and capacity control of polyaniline based organic cathodes: an ab initio study, *J. Power Sources*, 2016, **336**, 126-131.

42 Y. Chen, S. Manzhos, A computational study of lithium interaction with tetracyanoethylene (TCNE) and tetracyaniquinodimethane (TCNQ) molecules, *Phys. Chem. Chem. Phys.*, 2016, **18**, 1470-1477.

43 Sk Mahasin Alam, Y. Chen, S. Manzhos, Orbital order switching in molecular calculations using GGA functionals: qualitative errors in materials modeling for electrochemical power sources and how to fix them, *Chem. Phys. Lett.*, 2016, **659**, 270-276.

44 Y. Chen, J. Lueder, M. F. Ng, M. Sullivan, S. Manzhos, Polyaniline and CN-functionalized polyaniline as organic cathodes for lithium and sodium ion batteries: a combined molecular dynamics and Density Functional Tight Binding Study in solid state, *Phys. Chem. Chem. Phys.*, 2018, **20**, 232-237.

45 M. J. Frisch, G. W. Trucks, H. B. Schlegel, G. E. Scuseria, M. A. Robb, J. R. Cheeseman, G. Scalmani, V. Barone, G. A. Petersson, H. Nakatsuji, X. Li, M. Caricato, A. Marenich, J. Bloino, B. G. Janesko, R. Gomperts, B. Mennucci, H. P. Hratchian, J. V. Ortiz, A. F. Izmaylov, J. L. Sonnenberg, D. Williams-Young, F. Ding, F. Lipparini, F. Egidi, J. Goings, B. Peng, A. Petrone, T. Henderson, D. Ranasinghe, V. G. Zakrzewski, J. Gao, N. Rega, G. Zheng, W. Liang, M. Hada, M. Ehara, K. Toyota, R. Fukuda, J. Hasegawa, M. Ishida, T. Nakajima, Y. Honda, O. Kitao, H. Nakai, T. Vreven, K. Throssell, J. A. Montgomery, Jr., J. E. Peralta, F. Ogliaro, M. Bearpark, J. J. Heyd, E. Brothers, K. N. Kudin, V. N. Staroverov, T. Keith, R. Kobayashi, J. Normand, K. Raghavachari, A. Rendell, J. C. Burant, S. S. Iyengar, J. Tomasi, M. Cossi, J. M. Millam, M. Klene, C. Adamo, R. Cammi, J. W. Ochterski, R. L. Martin, K. Morokuma, O. Farkas, J. B. Foresman, and D. J. Fox, Gaussian 09, Revision D.3, Gaussian, Inc., Wallingford CT, 2016.

46 A. D. Becke, Density-functional thermochemistry. III. The role of exact exchange, *J. Chem. Phys.*, 1993, **98**, 5648-5652.

47 J. P. Perdew, K. Burke, M. Ernzerhof, Generalized gradient approximation made simple, *Phys. Rev. Lett.*, 1996, **77**, 3865-3868.

48 E. Sanville, S. D. Kenny, R. Smith, G. Henkelman, An improved grid-based algorithm for Bader charge allocation, *J. Comp. Chem.*, 2007, **28**, 899-908.

49 N. M. O'Boyle, A. L. Tenderholt, K. M. Langner, CCLIB: a library for package-independent computational chemistry algorithms, *J. Comp. Chem*., 2008, **29**, 839-845.

50 K. Momma, F. Izumi, VESTA 3 for three-dimensional visualization of crystal, volumetric and morphology data, *J. Appl. Crystallogr*., 2011, **44**, 1272-1276.

51 C. Kittel, Introduction to Solid State Physics, 8$^{th}$ Ed., John Wiley & Sons, Hoboken, NJ, 2005.

52 E. Kaxiras, Atomic and Electronic Structure of Solids, Cambridge University Press, Cambridge UK, 2003.

53 G. Otero, G. Biddau, C. Sanchez-Sanchez, R. Caillard, M. F. Lopez, C. Rogero, F. J. Palomares, N. Cabello, M. A. Basanta, J. Ortega, J. Mendez, A. M. Echavarren, R. Perez, B. Gomez-Lor, J. A. Martin-Gago, Fullerenes from aromatic precursors by surface-catalysed cyclodehydrogenation, *Nature*, 2018, **454**, 865-869.

54 S. Arabnejad, A. Pal, K. Yamashita, S. Manzhos, Effect of nuclear motion on charge transport in fullerenes: a combined Density Functional Tight Binding – Density Functional Theory investigation, *Frontiers in Energy Research*, 2019, **7**, 3.

55 A. Pal, K. W. Lai, Y. J. Chia, I. Jeon, Y. Matsuo, S. Manzhos, Comparative Density Functional Theory – Density Functional Tight Binding Study of fullerene derivatives: effects due to fullerene size, addends, and crystallinity on band structure, charge transport and optical properties, *Phys. Chem. Chem. Phys*., 2017, **19**, 28330-28343.

56 D. V. Konarev, R. N. Lyubovskaya, S. S. Khasanov, A. Otsuka, G. Saito, Formation and properties of $(C_{60}^-)_2$ dimers of fullerenes bonded by one and two σ-bonds in ionic complexes, *Mol. Cryst. Liq. Cryst.,* 2007, **468**, 227/[579]-237/[589].

57 J. C. Hummelen, B. Knight, J. Pavlovich, R. González, F. Wudl, Isolation of the heterofullerene $C_{59}N$ as its dimer $(C_{59}N)_2$, *Science*, 1995, **269**, 1554-1556.